*Article*

# Cell response on laser-patterned Ti/Zr/Ti and Ti/Cu/Ti multi-layer systems

Suzana Petrović [1,*], Nevena Božinović [1], Vladimir Rajić [1], Danijela Stanisavljević Ninković [2], Danilo Kisić[1] Milena Stevanović [2,3,4] and Emmanuel Stratakis[5]

1. Vinča Institute of Nuclear Sciences, University of Belgrade, P.O.Box 522, 11001 Belgrade, Serbia, office@vin.bg.ac.rs
2. Institute of Molecular Genetics and Genetic Engineering, University of Belgrade, Vojvode Stepe 444a, PO Box 152, 11042 Belgrade, Serbia
3. University of Belgrade, Faculty of Biology, Studentski trg 16, PO Box 43, Belgrade, 11000, Serbia
4. Serbian Academy of Sciences and Arts, Knez Mihailova 35, 11001 Belgrade, Serbia
5. Institute of Electronic Structure and Laser (IESL), Foundation for Research and Technology (FORTH), N. Plastira 100, Vassilika Vouton, 70013 Heraklion, Crete, Greece; stratak@iesl.forth.gr (E.S.)

\* Correspondence: spetro@vin.bg.ac.rs ; Tel.: +381 11 3408560

**Abstract:** Arranged patterns obtained by ultrafast laser processing on the surface of Ti/Cu/Ti/Si and Ti/Zr/Ti/Si thin film systems are reported. Two differently designed multilayer thin films, Ti/Cu/Ti/Si and Ti/Zr/Ti/Si, were deposited on silicon by the ion sputtering method. The bioactive surfaces on these systems involve the formation of laser-induced periodic surface structures (LIPSS) in each of the laser-written lines of mesh patterns on 5x5 mm areas. The formation of nano- and micro-patterns with an ultra-thin oxide film on the surfaces was used to observe the effects of morphology and proliferation of the MRC-5 cell culture line. To determine whether Ti-based thin films have a toxic effect on living cells, an MTT assay was performed. The relative cytotoxic effect as a percentage of surviving cells showed that there was no difference in cell number between the Ti-based thin films and the control cells. There was also no difference in the viability of the MRC-5 cells, except for the Ti/Cu/Ti/Si system, where there was a slight 10% decrease in cell viability.

## 1. Introduction

Implants are now becoming a standard therapy for repairing and replacing bone tissue, making daily activities much easier for patients. The implanted biomaterial should replace the missing bone while stimulating osteoconduction for bone re-growth. On the one hand, extensive research has been directed to improve the performance of implants in terms of biocompatibility and satisfactory mechanical properties [1,2]. New alloys have been developed, mainly based on titanium. On the other hand, surface modifications such as the adaptation of the chemical composition of the contact surface and the formation of specific morphological features at the micro- and nanometer level, further improve the performance of the implant. It has been shown that micro- and nanostructured surfaces are a good way to improve the bioactivity of biomaterials and osteogenic differentiation of cells [3].

Metal orthopedic implants were considered more promising due to their greater toughness and durability, but a large number of metals and their alloys, although they possess desirable bactericidal properties, can be very irritating and toxic in contact with the biological environment [4]. Orthopedic materials such as steel, titanium alloys (Ti-6Al-4V one of the most commonly used) and chromium alloys are widely used as orthopedic implants and devices, which have been adopted for dentistry and hard tissue reconstructive surgery. However, there are some rates of unwanted side effects as a tendency to corrosion, insufficient biocompatibility, and subsequent tissue irritation, which is limited their application [5,6]. The continuous development of materials for orthopedic implants requires knowledge of their mechanical and microstructural properties, as well as the development of new methods of surface modification in order to adapt the morphology to the best possible cellular response. For obtaining the large-scale incorporated layer (antibacterial or more biocompatible) in order to avoid the exfoliation from the substrate, a relatively new surface modification technology called friction stir processing (FSP) was developed to produce surface nanocomposite with enhanced mechanical properties in many alloy systems [7]. Also, one promising technique is selective laser melting (SLM) for manufacturing Ti-based alloys composed of dense blocks that have the advantages like low porosity and high yield strength, and the element segregation was suppressed significantly. This method has allowed the formation of strut-based latticed-beams structures with tunable Young's modulus, which is applicable to the complex bone structure [8].

Titanium (Ti)-based materials are good biomaterials due to their properties, such as biocompatibility, lightweight, high strength, and superior corrosion resistance [ 9,10]. For long-term bone replacements, titanium alloys should possess bone-mimicking mechanical properties, which can be obtained by the optimization of the geometry of the porous structure. Topological porous Ti-based implants also improve osseointegration and can be made patient-specific, which is essential for preventing their failure due to inadequate integration of the implant into the bone tissue [11]. Titanium alloys are suitable as biomaterials due to their superior biocompatibility, as the outer oxide layer is negatively charged at physiological pH and the soluble metal components cannot be released into the biological fluids [12,13]. A high concentration of β-stabilising elements in Ti-based alloys has been developed as a potential solution to the mismatch between the elastic modulus of the implant and the surrounding bone tissue. Zirconium (Zr) is the most common element added to Ti alloys, as it doesn't show any cytotoxic reaction when in contact with cells [14-16]. The crucial of the biomaterial's application is the formation of the implant-tissue interface during the interaction of the Ti alloy with the biological environment, which is largely defined by the surface characteristics of the alloy. Surface topography, composed of micro- and nanometer features, determines cell adhesion, morphology, proliferation, and differentiation, and therefore has a major impact on cell survival and implant-tissue interface properties [17]. Recently, the Ti-Nb-Zr-Ta alloy has attracted more and more attention due to the shape memory effect and superelastic behavior is associated with a reversible, thermoelastic transformation between the β phase and the α″ martensite phase of the alloy [18]. Ikarashi et al were implanting Ti, Zr, and Ti–Zr alloy into rats for 8 months, and no toxicological change in the body or organ weights as well as in hematological parameters. However, the tissue inflammatory responses to the Ti–Zr alloy was lower and the Ti–Zr alloy has better biocompatibility than other artificial surgical implants [19]. In preclinical studies, TiZr alloy, in combination with a SLActive® surface (Straumann's high-performance surface with extensive healing potential), is a suitable material for small-diameter implants. The results demonstrated good performance and tolerability with high implant stability and fast osseointegration [20,21].

Metallic orthopedic materials as a substitute for hard tissue in clinical use can release metallic ions and wear debris in artificial joints, and that are serious problems to be solved for safe application. Essentially, metals are non-biofunctional, their surfaces should be modified to improve corrosion resistance, wear resistance, and bone conductivity [1]. Surface modification of orthopedic material aims to enhance their performance in contact with the biological environment by forming the biointerface. Frequently used surface techniques are physicochemical methods that alter surface characteristics by physical texturing and/or chemical reactions including etching/oxidation, the addition of functional groups, deposition of functional thin films, ion or laser irradiation treatments and surface patterning by lithography [22]. Among a variety of surface modification techniques, the ultrafast laser modification technique has emerged as one of the best for forming ordered surface patterns, increasing surface roughness, changing composition, and functionalizing and forming a thin film [23]. This technique is suitable for the fabrication of bioactive surfaces, on stents, bone implants, and biofilms [24-26]. The energy of laser radiation can be focused on the surface and modify only selected areas, triggering the transformation of hardening, surface melting, surface oxidation, and/or surface alloying [27].

Laser processing of biomaterials is a particularly important opportunity for the application of multiple pulsed laser effects that produce laser-induced periodic surface structures (LIPSS). When the periodicity of LIPSS is nearly equal to the laser wavelength and oriented normal to the polarization vector of the laser radiation, they are referred to as low spatial frequency LIPSS (LSFL) [28,29]. Another type of periodic structures, called high spatial frequency LIPSS (HSFL), has much lower spatial periodicity but is aligned parallel to the polarization vector [30]. Advances in the field of ultrafast laser processing based on ablation of materials are very important due to the generation of 2D surface and 3D volume micro- and nanopatterns. The one of main advantages of ultrafast laser pulses for ablating of materials is capable of producing extremely high peak radiation intensities (>$10^{11}$ W/cm$^2$), generating free electrons through a multiphoton absorption mechanism and band-gap tunnelling. A high density of excited electrons can be created locally in a small focal volume due to nonlinear absorption, triggering avalanche ionization that supports plasma formation. Consequent to that is the fabrication of a smoother cavity wall with negligible thermal damage [31,32]. Recently, ultrafast laser surface modification has evolved into a unique method that enables the fabrication of a bioactive surface with the formation of the desired oxide and alloy, the generation of nano/microtextures, and the modification of the wettability of the surface [33-36]. The optimal surface structure can be achieved by different surface patterns (physical shape and size, roughness, regular or irregular series of patterns, wettability, surface energy, etc.), with the appropriate surface composition (oxides that prevent the release of toxic metal ions), in order to stimulate cell growth.

The main topic of this work is to investigate the possibility of obtaining bioactive surfaces on Ti/Zr/Ti/Si and Ti/Cu/Ti/Si systems, with adapted morphology and wettability for cell integration. The reason for choosing the Ti/Zr/Ti/Si and Ti/Cu/Ti/Si systems is reflected in the fact that titanium alloys with Zr have satisfactory mechanical strength and good corrosion resistance in biological fluids, and their biocompatibility is better than commercially pure Ti for implants [37]. On the other hand, copper as an essential trace element required for human health, is one of the most promising alloying elements for clinical applications due to its low toxicity and high cytocompatibility. Recently, several studies have reported that titanium alloys with Cu have excellent antibacterial function [38]. It was reported that Ti-(1 wt.% and 5 wt.%) Cu alloys showed antibacterial properties with an antibacterial rate of about 30% in comparison with pure titanium [39]. Precisely, Zr and Cu are in small amounts just below the surface, so that after laser modification they would be in contact with the biological environment. The research focused on the formation of laser-printed templates with micro- and nanometer dimensions and inherent wettability suitable for the positioning of cells to enable their proliferation and growth. The systematization of the obtained results included a comparison of the changes in the Ti/Zr/Ti/Si and Ti/Cu/Ti/Si systems deposited and modified under the same experimental conditions. Specifically, the results were intended to show the influence of the type of alloy component added to the Ti matrix on the properties of the bioactive surface. The MTT test will show the degree of cell survival on different Ti/Zr/Ti/Si and Ti/Cu/Ti/Si systems with almost the same surface topography.

## 2. Materials and Methods

*Thin film deposition*. The thin multilayer films of Ti/Cu/Ti/ and Ti/Zr//Ti were deposited on the single crystal (100) of n-type silicon in Balzers Sputtron II system by ion sputtering of pure Ti, Cu, and Zr with 1.3 keV argon ions. Before chamber assembly, the surfaces of the Si substrates were cleaned with ethanol and dilute hydrofluoric acid (HF) to remove impurities and native Si oxides. The base pressure in the chamber was $5.5 \times 10^{-4}$ Pa. Targets were sputtered with a constant current of 0.7 A and voltage of 0.5 kV for Ti and corresponding current and voltage values of 0.5 A and 1.2 kV for deposition of Cu and Zr, respectively. The deposition rates were constant at about 0.13 nm s$^{-1}$ for Ti and 0.25 nm s$^{-1}$ for Cu and Zr. The total thickness of the deposited Ti/Cu/Ti and Ti/Zr//Ti thin films was 300 nm, with the subsurface Cu and Zr layers having a thickness of 10 nm and the top Ti layer also having a thickness of 10 nm

*Ultra-fast laser modification*. Laser processing of these specific Ti/Zr/Ti/Si and Ti/Cu//Ti/Si systems was performed using the Yb: KGW laser source Pharos SP from Light Conversion. The surfaces of the thin films were irradiated with focused, linear p-polarized pulses with the following characteristics: repetition rate of 5 kHz, pulse duration equal to 160 fs, and central wavelength of 1026 nm. The joulemeter is used to measure the energy deposited on the surface of the sample before real irradiation. In each irradiation, the pulse energy was assumed to be constant, as the deviation of the pulse energy was less than 1%. The laser beam pulses were guided through proper optical mirrors and an optical lens and focused on the sample surface with an f = 200 mm plano-convex lens. The beam diameter was characterized by a CCD camera close to the focal plane and was estimated at around ~ 60 μm for the Gaussian beam, with corresponding spot size at focus $2.8 \times 10^{-5}$ cm$^2$ (spot diameter 30 μm). Irradiations were performed with a fluence of $0.110 \pm 0.005$ J cm$^{-2}$ (calculated as $\Phi_0 = 2E_P/\pi\omega_0^2$ with $E_P$ the pulse energy and $\omega_0$ the beam radius at $1/e^2$). The samples were laser processed in the open air and mounted on a motorized computer-controlled X-Y-Z translation perpendicular to the laser beam. The irradiation strategy included formation mesh at the surface of sample 5 x 5 mm composed of laser-printed lines at a scanning speed of 15 mm s$^{-1}$ (20 pulses per spot). Firstly, the lines of 5 mm length are written at a distance of 75 μm, then the sample is turned for 90° and the procedure is repeated so that finally a mesh of laser-printed lines is obtained.

*Morphological characterization*. The surface morphology and microstructure of the Ti/Zr/Ti/Si and Ti/Cu/Ti/Si thin films before and after laser processing were analyzed by a FESEM, FEI SCIOS 2 Dual Beam scanning electron microscope (SEM) equipped with energy-dispersive X-ray spectroscope (EDS) at 10 kV of acceleration voltage.

*Surface free energy (SFE) determination using the sessile drop method*. The wettability measurements were performed on a home-made device equipped with a polychromatic LED lamp as a counter light source (manufacturer YINYANG) and USB digital CCD camera (König Electronic USB microscope, model CMP-usbmicro10). The working distance between lens and sample holder was 15 mm, with a magnification of 36x for the diagonal field of view of 8 mm, while the scale bar length is $100 \pm 1$ pixels per milimeter.. The basic data in this study were the contact angle (CA - θ), which indicates the degree of wetting as the contact between the surface and the liquid. The CA with the surface of the samples on untreated and laser-treated thin films was measured at a room temperature of 23° C and a relative humidity of 53-56% with three different liquid drops (water, ethylene glycol, diiodomethane). Droplets were laid on the surfaces of samples using micropipette, and the droplet volume was set to 3 μl. To obtain the average value, contact angle measurements were taken from at least six separate regions of the same sample. Image analysis was performed with public ImageJ software [40].. The SFE and its polar ($\gamma_P$) and dispersive components ($\gamma_d$) of the metallic surfaces were calculated using the Good-van-Oss model based on the contact angle data [41].

*Study of cellular response*. MRC-5 cells (ATCC®, CCL-171TM) were maintained in Dulbecco's Modified Eagle's Medium (DMEM) supplemented with 10% fetal bovine serum (FBS), 4500 mg/L glucose, 2 mmol/L L-glutamine, and antibiotic/antimycotic (all from Invitrogen™, NY, USA) at 37 °C and 10% CO$_2$. Cells were observed and images were acquired using DM IL LED inverted microscope (Leica Microsystems, Wetzlar, Germany). Two pieces of each thin film (Ti, Ti/Cu/Ti, and Ti/Zr/Ti) with a size of 10 mm × 10 mm were incubated with DMEM culture medium under sterile conditions at 37 °C and 10%

CO₂ for 10 days. After incubation, the thin film was removed, and the remaining medium was considered 100% extract. Diluted extract (50% and 15%) was used for experiments.

The indirect method was used to measure the cytotoxicity of the film. MRC-5 cells were grown at a density of 10 000 cells on a 96-well plate and cultured the next day with 100%, 50%, and 15% of the extraction medium obtained from the incubated films with DMEM for 10 days at 37 °C. After 24 hours of incubation, cell viability was determined using the MTT (3-(4,5-dimethylthiazol-2-yl)-2,5-diphenyltetrazolium bromide) assay according to the manufacturer's protocol. Absorbance was measured at 540 nm using an Infinite 200 pro plate reader (Tecan, Austria). Two controls were used: MRC-5 cells treated with DMEM and incubated at 37°C for 10 days without films (referred to as Blank), and MRC-5 cells treated with fresh DMEM (referred to as C). As a positive control for cytotoxicity, we use MRC-5 cells treated with pyrimethanil (denoted as P), whose cytotoxic effect on MRC-5 cells has been previously demonstrated [42]. Statistical analyses were performed using SPSS statistical software (version 20). The results obtained are means ± SEM from four independent experiments. Statistical analyses were performed using Student's t-test, and a p-value ≤.05 was considered significant, with $p \leq .05$ denoted as *.

## 3. Results and Discussion

### 3.1. Morphological characterization of Ti/Zr/Ti/Si and Ti/Cu/Ti/Si thin film systems

Femtosecond multipulse laser exposure to Ti/Zr/Ti/Si and Ti/Cu/Ti/Si thin film systems was used to create a regular mesh of laser-printed lines forming micrometre-sized morphological features. To achieve a suitable position of the cells on the surface, the diameter of the laser beam was set to about 30 μm, while the distance between the lines was about 75 μm [43]. The morphological characteristics of the Ti/Zr/Ti/Si thin film system after laser processing were analyzed by SEM method (Figure 1). Laser-induced periodic surface structures (LIPSS) aligned normal to the polarization vector of the applied laser beam are formed in each laser-printed line, which classifies them as low spatial frequency LIPSS (LSFL) [28,29]. The formation of LSFL was accompanied by some degree of material removal, which contributed to their well-defined shape, especially at the intersections (Figure 1). Nanoparticles of size 70 - 150 nm have accumulated at the edges of the lines, which is due to the pronounce ablation during the formation of the periodic structure in the central part. In the crossing region, the LSFLs are narrow and densely distributed with a spatial periodicity of about 700 nm. On the other hand, the spatial periodicity of the ripple structure in the lines was about 1 μm (very close to the laser wavelength), so the LSFLs are more widely and less frequently distributed (Figure 1). The square inner regions between the lines, which are not processed by the laser, are partially covered with nanoparticles.

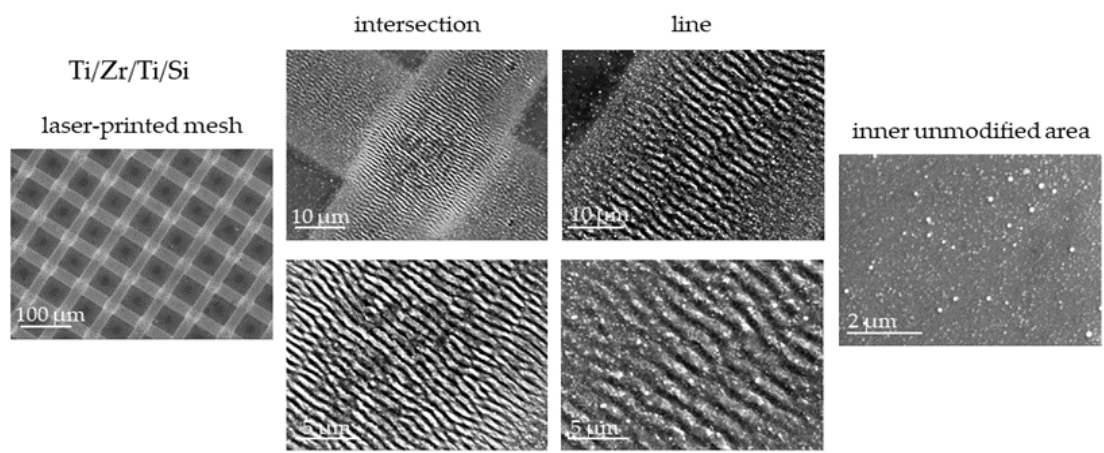

**Figure 1.** SEM micrographs after laser-printed line mesh on the surface of Ti/Zr/Ti/Si thin film system with morphological details at an intersection, line and inner unmodified areas.

The Ti/Cu/Ti/Si system was modified under the same experimental laser processing conditions as the Ti/Zr/Ti/Si system. In general, the observed morphological features are similar to the previous case (Figures 1 and 2), except for some details: (i) the width of the laser-printed line is narrower, which is a consequence of the less pronounced ablation of the material, (ii) the periodic structure (LSFL) in the line is not well defined due to the occurrence of hydrodynamic effects in the form of partial surface melting, (iii) among the unclear LSFL ripples, the additional appearance of high spatial frequency LIPSS (HSFL) with a periodicity below 200 nm and an orientation parallel to the polarization is observed, and (iv) at the edges of the lines, there is no larger number of nanoparticles, which is probably due to the effect of surface melting (Figure 2). It is interesting to note that the Cu sublayer with a thickness of 10 nm significantly reduces the ablation effect, which is due to the thermophysical properties of Cu and Zr [44,45]. The higher mobility of Cu atoms, as well as the higher thermal conductivity of copper, may contribute to an easier redistribution of the absorbed laser radiation energy in the Ti/Cu/Ti/Si system.

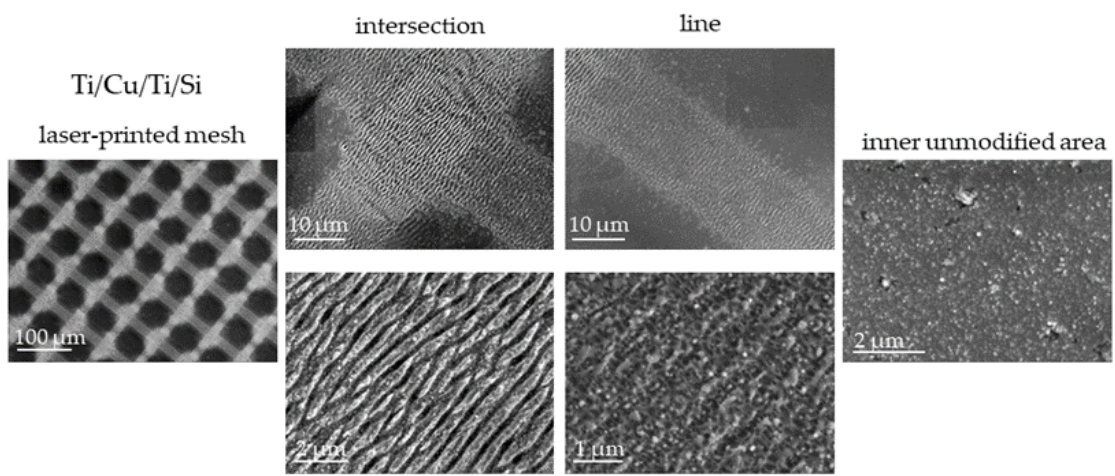

**Figure 2.** SEM micrographs after laser-printed line mesh on the surface of Ti/Cu/Ti/Si thin film system with morphological details at an intersection, line and inner unmodified areas.

Mapping of the laser-modified surface for both systems studied was performed using EDS and is shown in Figure 3. The distribution of the thin film components (Ti, Zr or Cu, Si and O) follows the pattern of a laser-printed mesh, which superficially appears very similar for both systems.

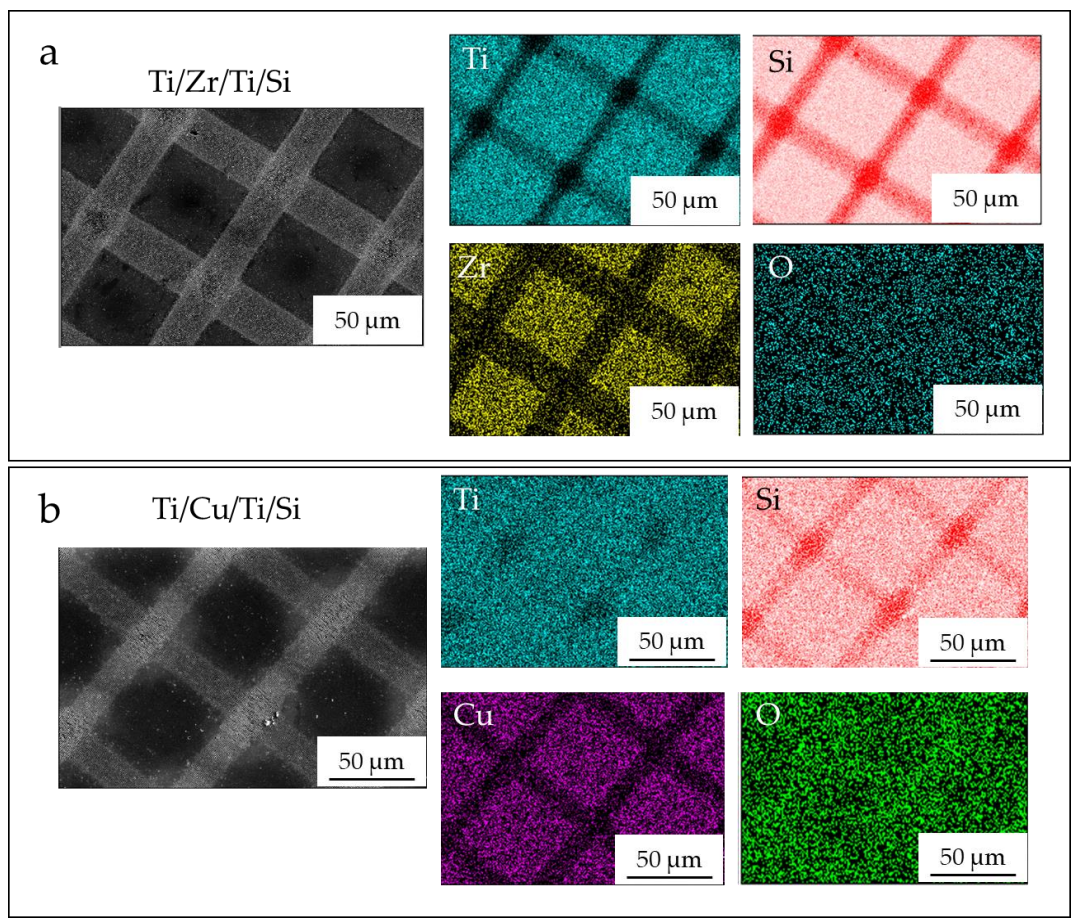

**Figure 3.** EDS mapping of the laser-modified surface for (a) Ti/Zr/Ti/Si and (b) Ti/Cu/Ti/Si thin film systems, with corresponding elemental maps.

In the individual elemental maps for Ti and Si in both systems, it can be seen that the ablation is more intense in the Ti/Zr/Ti/Si system, which is due to the presence of a larger number of Si atoms at the positions of the laser-printed lines and, in particular, at the intersection positions (Fig. 3 a and b). In the Ti/Cu/Ti/ Si system on the elemental map for Ti, deviations from the uniform distribution of Ti were observed only at the intersection points (Fig. 3 b). Comparing the distribution of Zr and Cu on the corresponding elemental maps (Fig. 3 a and b), it was found that the ablation of Zr is much more pronounced in the laser-printed lines, which is evident in a larger area where the concentration of the Zr component is reduced. However, the ablation is not complete at the locations of the laser-printed lines because the ablated material is redeposited in the form of nanoparticles. There was additional ablation in the intersections, so the presence of Zr or Cu components is not expected. The oxygen is mostly uniformly distributed over the surface in both sample types and originates from laser-induced surface oxidation or atmospheric contamination. In the Ti/Cu/Ti/Si samples, a slightly higher oxygen concentration is observed at the positions of the laser-printed lines, suggesting that lower ablation of the material is associated with more intense surface oxidation of Ti and Cu species.

A comparison was also made of the concentrations at specific points corresponding to the unmodified area (as deposited), the inner unmodified area (centre), the laser-printed lines, and the intersection areas, as shown in Table 1 and Table 2. In the Ti/Zr/Ti/Si system, significantly higher material removal (more than 30%) was obtained, and no Cu or Zr sublayer components are present at the intersections. The higher concentration of Ti after modification was achieved by norming the concentration of the components in the analyzed information volume of the sample. During laser irradiation, slightly more Cu remained in the lines than Zr. The results indicate that oxygen from the atmosphere more readily oxidized the surface where copper was present. Taking into account current and previous studies of a similar Ti/Zr multilayer system, it can be concluded that a larger fraction of the laser pulse energy delivered to the system induced intense ablation (removal of between 100 and 200 nm thin film), while a smaller fraction of the absorbed energy induced changes in the composition coinciding with a change in lattice temperature (in-depth about 50 nm) based on numerical simulation of the temperature profile [36,46,47].

**Table 1.** Concentrations (weight %) of Ti, Si, Zr and O species in the Ti/Zr/Ti/Si sample at the positions of the unmodified part (as-deposited), the inner unmodified part (centre), laser-printed lines and intersection regions.

| **Ti/Zr/Ti/Si** | **Ti** | **Si** | **Zr** | **O** |
|---|---|---|---|---|
| As-deposited | 77.03 | 16.21 | 5.81 | 0.95 |
| centre | 76.60 | 16.75 | 5.67 | 0.98 |
| line | 49.37 | 47.88 | 0.74 | 2.01 |
| intersection | 43.91 | 53.63 | / | 2.46 |

**Table 2.** Concentrations (at %) of Ti, Si, Cu and O species in the Ti/Cu/Ti/Si sample at the positions of the unmodified part (as-deposited), the inner unmodified part (centre), laser-printed lines and intersection regions.

| **Ti/Cu/Ti/Si** | **Ti** | **Si** | **Cu** | **O** |
|---|---|---|---|---|
| As-deposited | 72.92 | 14.85 | 10.91 | 1.32 |
| centre | 73.09 | 15.04 | 10.51 | 1.36 |
| line | 74.37 | 17.92 | 5.08 | 2.63 |
| intersection | 74.62 | 22.35 | / | 3.03 |

Surface free energy (SFE) can be considered a quantity that characterizes the affinity of a surface for other materials (e.g., liquids). Contact angle measurements were used to study the Ti-based thin films, as-deposited and after laser treatment using deionized water, ethylene glycol, and diiodomethane as reference fluids. The SFE ($\gamma_s$) contains dispersive $\gamma_s^d$ (Lifshitz-Van der Waals) and polar $\gamma_s^p$ (Lewis acid $\gamma_s^+$ and base $\gamma_s^-$) components related by the equation [48]. Based on the results in Table 3, where the average contact angle values are given, it can be observed that all samples had higher contact angles after laser irradiation and the best possible wettability for diiodomethane was achieved.

The dispersive component was much more pronounced in all samples than the polar component, which did not change after irradiation. This proportion of surface energy components indicates that the sample surfaces have a stronger preference for nonpolar organic liquids than for water. Surface hydrophobicity is primarily due to the self-association of water molecules upon contact with a nonpolar surface. Accordingly, surface hydrophobicity (water un-wettability) refers to the poor solubility of nonpolar solutes such as hydrocarbons in water according to the rule of equality [49]. From Table 3, it can be seen that the value of total SFE slightly increases for the Ti/Zr/Ti/Si layer, which means that the laser irradiation did not significantly change the dispersive character of the surfaces, even though the removal was more severe compared to the Ti/Cu/Ti/Si layer.

**Table 3.** Surface energies SFE ($\gamma_s$) and its components, dispersive contact angles of the films wetted with deionized water and diiodomethane.

| Specimen | Contact angle [°] | | | Surface free energy [mJ/m²] | | |
|---|---|---|---|---|---|---|
| | Deionized water | Diiodmethane | Ethylene glycol | Polar part ($\gamma_s^p$) | Nonpolar part ($\gamma_s^d$) | Total ($\gamma_s$) |
| Ti/Cu/Ti as-prepared | 84 ± 3 | 50 ± 3 | 54 ± 2 | 1,7 | 33,7 | 35,4 |
| Ti/Cu/Ti laser modified | 86 ± 4 | 51 ± 3 | 63 ± 3 | 0,6 | 34,3 | 34,9 |
| Ti/Zr/Ti as-prepared | 82 ± 2 | 37 ± 2 | 58 ± 3 | 3,3 | 40,1 | 43,4 |
| Ti/Zr/Tu laser modified | 86 ± 3 | 40 ± 2 | 70 ± 3 | 6,3 | 40,1 | 46,4 |

Therefore, Table 4 explains in more detail how the surface changes after laser processing. While the Lewis acid component dominates in the laser-modified Ti/Zr/Ti/Si, the Lewis base component predominates in both irradiated samples, indicating that chemical species containing only one electron pair became dominant after laser treatment. This could be the result of the increase in surface oxygen concentration and the formation of oxygen-containing species, which was confirmed by the method SEM (EDS)

**Table 4.** Calculated values of the Lewis acid $\gamma_s^+$ and Lewis base $\gamma_s^-$ components.

| Sample | $\gamma_s^+$ [mJ/m²] | $\gamma_s^-$ [mJ/m²] |
|---|---|---|
| Ti/Cu/Ti as-prepared | 0.77 ± 0.03 | 0.92 ± 0.03 |
| Ti/Cu/Ti laser modified | 0,02 ± 0,04 | 4,82 ± 0,03 |
| Ti/Zr/Ti as-prepared | 0,29 ± 0,02 | 9,14 ± 0,02 |
| Ti/Zr/Tu laser modified | 1,30 ± 0,03 | 7,53 ± 0,02 |

The wettability of Ti/Zr/Ti/Si and Ti/Cu/Ti/Si surfaces was investigated by CA measurements before and after laser treatment. Using Image J Drop Analysis software, the wetting angles were calculated with deionized water, diiodomethane and ethylene glycol and shown in the images in Fig. 4 and Fig. 5. It can be seen that the surfaces of both samples are hydrophobic after laser irradiation, taking into account that the contact angles with water are slightly larger than before. Following Berg's limits based on measurements of hydrophobic forces [50,51], it can be concluded that the Ti/Zr/Ti/Si and Ti/Cu/Ti/Si thin films are supported by hydrophobic forces and are less water wettable. Due to the energetically beneficial displacement of solutes from solution into the interphase between solid and solute phases, hydrophobic surfaces facilitate the adsorption of various surfactants and proteins from water [52]. When important measures of biological activity responsive to interfacial phenomena are related to the water adhesion tension of the contacting surfaces, it becomes clear that the physicochemical properties of the interfacial water have a strong influence on the biological response to materials.

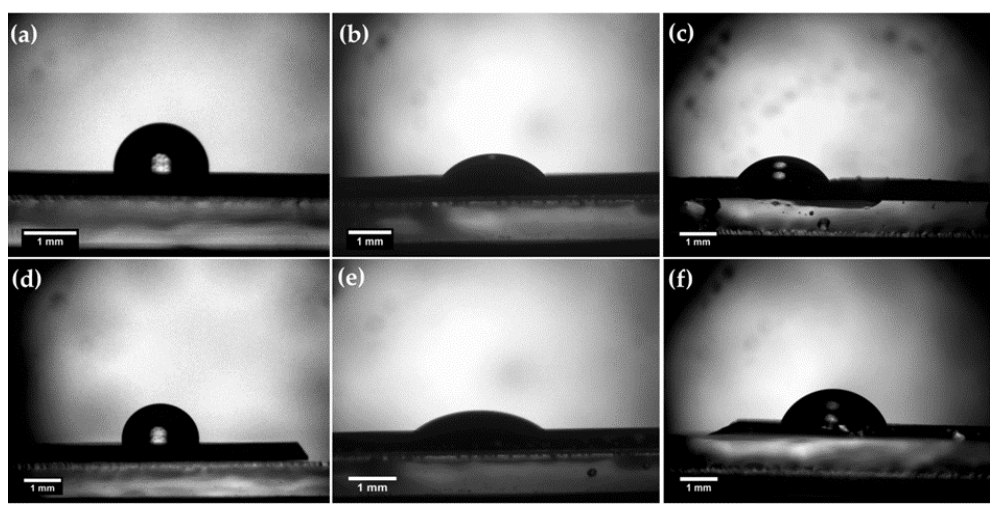

**Figure 4.** Static behaviour of a 3 μm drop liquids on the metallic surfaces of as-prepared (a), (b), (c) Ti/Cu/Ti/Si thin film wetted with deionized water, diiodmethane and ethylene glycol respectively; and (d), (e), (f) Ti/Zr/Ti/Si thin film wetted in the same order. Ti is the uppermost layer.

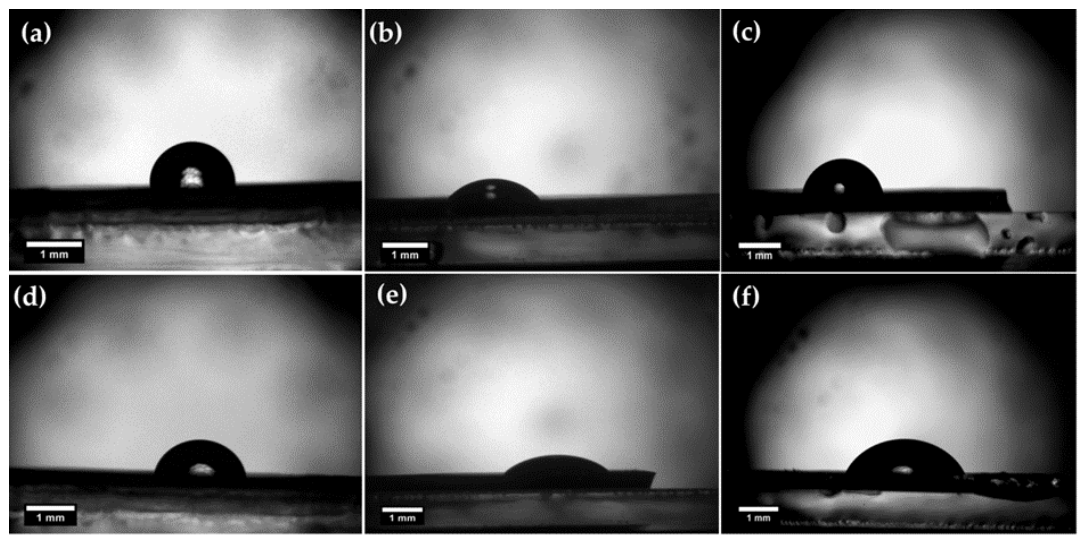

**Figure 5.** Static behaviour of a 3 μm drop liquids on the metallic surfaces after femtosecond laser irradiation for (a), (b), (c) Ti/Cu/Ti/Si thin film wetted with deionized water, diiodmethane and ethylene glycol respectively; and (d), (e), (f) Ti/Zr/Ti/Si thin film wetted in the same order.

Thin Ti-based films (Ti/Cu/Ti and Ti/Zr/Ti) prepared by ion sputter deposition techniques and subsequently modified with a femtosecond laser pulse were subjected to cytotoxicity testing (MTT) to gain the most realistic insight into the behaviour of the tested materials in the human body. The viability of MRC-5 cells in contact with the tested materials after 24 hours is shown in Figure 6a. The lowest mean value of viability of about 85% was observed on the laser-irradiated surface of Ti/Cu/Ti/Si in a medium diluted to 50%. While MRC-5 cells on the surface of Ti/Zr/Ti/Si showed a higher mean viability value of over 100% compared to a control sample. The following order represents the ranking of the materials based on the percentage of MRC-5 cells surviving over the selected time period: 100% Ti/Zr/Ti > 15% Ti/Zr/Ti > 50% Ti/Zr/Ti > 100% Ti/Cu/Ti > 15% Ti/Cu/Ti > 50% Ti/Cu/Ti. However, after 24 hours, the degree of survival of cells in contact with pure Ti materials is roughly similar to or greater than the degree of survival of cells in contact with the control material. The typical value of viability after 24 hours is between 102 and 107%.

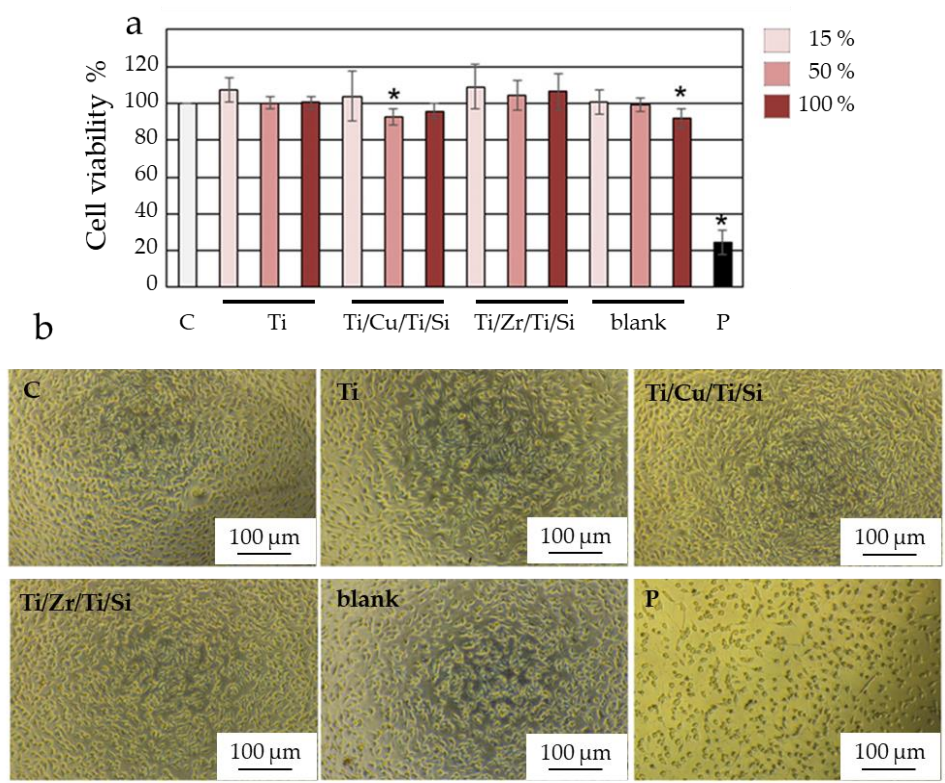

**Figure 6.** Cytotoxic assay of Titanium films on MRC5 cells. A) Histogram representation of the effect of Titanium films on MRC-5 viability, quantified upon four independent experiments with technical replicates. The relative cytotoxic effect of Titanium films was calculated as a percentage of survived cells upon treatments with three different concentrations of an extraction medium obtained from the incubated Titanium films with DMEM for 10 days at 37 °C compared to cells treated with fresh DMEM (designed as C) that was set as 100%. MRC-5 cells treated with DMEM incubated for 10 days at 37 °C without films (designed as Blank) were used as an additional control. As a positive control for cytotoxicity, we use MRC-5 cells treated with pyrimethanil (designed as P). Results were presented as the means ± SEM of four independent experiments. p values were calculated using Student's t-test, *$p \leq 0.05$. B) MRC5 cells 24 h after treatment with an extraction medium obtained from the incubated Titanium films with DMEM for 10 days at 37 °C. Cells treated with fresh DMEM (designed as C) were used as control. As a positive control for cytotoxicity was used MRC-5 cells were treated with pyrimethanil (designed as P). Cells were monitored under the microscope at a magnification of 20 x.

It can be concluded that the mean viability of MRC-5 cells in indirect contact with all tested materials exceeds the mean viability of cells in contact with the control material. On the surface of all materials tested, MRC-5 cells had a typical triangular, elongated spindle, and star form with conspicuous cytoplasmic appendages (Figure67b). The results demonstrated that both tested groups of materials have appropriate biocompatibility and are safe for use in the human body. The results of the conducted MTT test showed that the materials obtained by the method of cathodic sputtering and modified by laser rays are not cytotoxic.

It is common for laser modification of metallic materials to lead to the formation of nanoparticles as a result of the condensation of the ablated material. In particular, the higher number of formed nanoparticles with a wide range of dimensions occurs during the irradiation by femtosecond laser pulses due to the weak pronounced hydrodynamic effects on the sample surface [28,32,33]. Metallic nanoparticles showed a wide range of positive effects on stem cell differentiation and have a number of applications in regenerative medicine. However, metallic nanoparticles also have negative effects on stem cells in a number of ways. One of the common mechanisms is through the production of reactive oxygen species (ROS), which can cause oxidative stress and damage to cellular components such as proteins, lipids, and DNA. Nanoparticles can also disrupt cellular membranes and alter cellular signalling pathways, leading to changes in cell behaviour and function. In addition, nanoparticles can accumulate in cells and tissues, potentially leading to toxicity over time. The specific mechanisms by which nanoparticles affect cells can vary depending on the properties of the nanoparticles themselves, as well as the type of cell and tissue they are interacting with [53, 54]. The negative effects of nanoparticle-cell interactions depend on several factors such as the size, shape, surface charge, and surface chemistry of the nanoparticles. The interaction between the nanoparticles and the cell membrane can lead to internalization of the nanoparticles by the cell. The process can occur through several pathways such as endocytosis, phagocytosis, or direct penetration of the cell membrane [55]. Nanoparticles of $TiO_2$ showed a toxic effect in MSCs in a size-dependent manner, which was highlighted by low cell migration, lack of cell membrane integrity, and suppression of osteogenic differentiation [56]. Also, nanotubes of $TiO_2$ larger than 50 nm showed a drastic decrease in the proliferation and differentiation of MSCs [56]. Nanoparticles on the titanium implant can affect the biological function of the surrounding cells in two ways, firstly by affecting protein adsorption around cells and secondly by penetrating directly by cells. The degree of this effect in both cases depends on particle size and nanotopography. Results confirmed that the $TiO_2$ nanoparticles (15-200 nm) had a cytotoxic effect on MSC depending on their concentration. This finding could be interpreted that highly concentrated $TiO_2$ nanoparticles produce oxidative stress as well as reactive oxygen species, leading to cell death. Furthermore, it was observed that the growth and division of MSC were negatively affected in a size-dependent manner by the $TiO_2$ nanoparticles potentially via a mechanism causing intracellular acidity, thus inhibiting cell proliferation [56]. In our experiment, no decrease in the number of cells was observed, probably due to additional preparation of samples before cell cultivation or the number and dimension of laser-produced nanoparticles being below a critical value that would negatively affect the number of surviving cells.

It was confirmed that laser-processed Zr-based materials are capable of modulating cell attachment, alignment and proliferation and that hMSCs undergo a faster and greater osteogenic differentiation, even when cultured with no supplemented media [57]. Jiao et al. found that MG63 osteoblast-like cells on the groove-textured surface of Ti-Zr alloy exhibited higher viability and better adhesion compared to those on the original untextured surfaces [58]. They suggested that the higher surface roughness, the increased presence of metallic oxides and the enhanced hydrophilicity of the groove-textured sample were the main contributors to its improved cytocompatibility. Busuioc et al. have done a biocompatibility evaluation on MSCs cells via the cell proliferation assay (MTT) for Ti-Zr substrates, which indicates a lack of cytotoxicity [59]. Cell proliferation is favoured and cells can survive in the long term. The fluorescence microscopy images show normal development of the cells, which maintain their shape and have an unchanged metabolism. In accordance with these and our current results regarding the laser-processed Ti/Zr multilayer thin films, from a biological point of view, these systems are shown improved ability to sustain cell survival and promote cell proliferation, as well as their orientation along laser-induced periodic surface structures.

## 4. Conclusions

The bioactive surface of Ti-based thin films, which improves cell viability, was achieved by depositing Ti/Zr/Ti/Si and Ti/Cu/Ti/Si systems and ultrafast laser processing. The micro- and nanometer morphological features in the form of arranged patterns were obtained by laser-printed lines incorporated into a regular mesh of 5 mm × 5 mm area. Laser-induced periodic surface structure (LIPSS) is formed in the laser-printed lines, which is oriented normal to the polarization of the laser beam and defined as low spatial frequency LIPSS (LSFL). The different width of the lines (wider for the Ti/Zr/Ti/Si system) is a consequence of the different degree of material removal during laser irradiation. Analysis of the elemental composition at specific points on the modified surface, as well as mapping of the surface to determine the distribution of components, led to the conclusion that more intense material removal was obtained for the Ti/Zr/Ti/Si system, whereas for the Ti/Cu/Ti/Si, the surface oxidation was somewhat more pronounced.

The biocompatibility of the laser-textured Ti/Zr/Ti/Si and Ti/Cu/Ti/Si systems was determined based on the viability of MRC-5 cells in indirect contact with the samples. The results clearly show that both tested thin film systems have adequate biocompatibility and are safe for use in the human body. The results of the MTT test show that these materials are not cytotoxic, and the typical value of viability after 24 hours is more than 100%. Bioactivation of these specific multilayer systems with laser surface texturing and tailoring of surface morphology and wettability could be useful for tissue engineering and implant applications.

**Author Contributions:** For research articles with several authors, a short paragraph specifying their contributions must be provided. The following statements should be used "Conceptualization, S.P., M.S. and E.S.; methodology, N.B., V.R., D.K. and D.S.N.; validation, S.P, M.S. and E.S.; formal analysis, N.B. V.R. D.S.N. and D.K.; investigation, S.P. and N.B.; writing—original draft preparation, S.P., D.S.N. and N.B.; writing—review and editing, S.P; supervision, S.P.; All authors have read and agreed to the published version of the manuscript."

**Funding:** This research was funded by the EU-H2020 research and innovation program under grant agreement NFFA-Europe-Pilot (n. 101007417) having benefitted from the access provided by Foundation for Research and Technology Hellas (FORTH) access provider (Institute of Electronic Structure and Lasers i.e., Institution) in Heraklion, Crete, Greece within the framework of the NFFA-Europe Transnational Access Activity.

**Acknowledgments:** This work was supported by the Ministry of Science, Technological Development and Innovation of the Republic of Serbia (projects No. 451-03-47/2023-01/ 200017 and No. 451-03-47/2023-01/ 200042). We also acknowledge the support from European Community, COST Action. Project No. CA21159 (PhoBios).

**Conflicts of Interest:** "The funders had no role in the design of the study; in the collection, analyses, or interpretation of data; in the writing of the manuscript; or in the decision to publish the results".